\documentclass[11pt,twocolumn]{article}
\usepackage{amsmath}
\usepackage{amssymb}
\usepackage{epsfig}

\begin{document}

\setlength{\baselineskip}{12pt}

\title{Coalition Formation and Combinatorial Auctions; Applications to Self-organization and Self-management in Utility Computing}
\author{Dan C. Marinescu  and Ashkan Paya \\
Computer Science Division \\
Department of Electrical Engineering and Computer Science \\
University of Central Florida, Orlando, FL 32816, USA \\
Email: [dcm, apaya]@cs.ucf.edu \\\\
John P. Morrison \\
Computer Science Department \\
University College Cork. Cork, Ireland \\
Email: j.morrison@cs.ucc.ie}

\maketitle

\begin{abstract}
In this paper we propose  a two-stage protocol for resource management in a hierarchically organized cloud. The first stage exploits spatial locality for the formation of coalitions of supply agents; the second stage, a combinatorial auction, is  based on a modified proxy-based clock algorithm and has two phases, a clock phase and a proxy phase. The clock phase supports price discovery; in the second phase a proxy conducts multiple rounds of a combinatorial auction for the package of services requested by each client. The protocol strikes a balance between low-cost services for cloud clients and a decent profit for the service providers.  We also report the results of an empirical investigation of the combinatorial auction stage of the protocol.
\end{abstract}


\section{Introduction and Motivation}
\label{IntroductionAndMotivation}
\medskip

Nowadays large farms of computing and storage servers are assembled to support several cloud delivery models including Software as a Service (SaaS), Platform as a Service (PaaS), and  Infrastructure as a Service (IaaS). In such systems users pay only for computing resources they use, similarly to other utilities such as electricity and water.

Computer clouds raise the question of how far we can push the limits of  composability of computing and communication systems, while still being able to support effective policies for resource management and their implementation mechanisms. The software, the glue allowing us to build increasingly more complex systems, consists of more and more layers thus,  the challenge of controlling large-scale systems is amplified.

Control theory tells us that accurate state information and a tight feedback loop are the critical elements for effective control of a system. In a hierarchical organization the quality of state information degrades as we move from the bottom to the top; {\it only local information about the state of a server is by definition accurate.} Moreover, this information is volatile, it must be acted upon promptly because the state changes rapidly.  Our recent results \cite{Marinescu15} confirm that hierarchical control has considerably larger overhead than a simple economic model for cloud resource management. The communication complexity of hierarchical control based on monitoring is more than two orders of magnitude higher and consumes a significant fraction of the available bandwidth at all levels of the interconnection network.

Existing solutions for cloud  resource management are neither effective nor scalable and they require detailed models of the system and accurate information about the state of individual servers. The  polices for cloud resource management must  support: (i) admission  control; (ii) capacity allocation; (iii) load balancing; (iv) energy optimization, and (v)  quality of service (QoS) \cite{Marinescu13}. Some of the existing and future challenges for cloud resource management which affect these policies are:

\smallskip

\noindent -1. The cloud infrastructure is increasingly more heterogeneous;  servers with different configurations of multi-core processors, GPUs, FPGAs, and data flow engines are expected to become elements of the cloud computing landscape.  

\smallskip

\noindent -2. The spectrum of cloud services and cloud applications widens. For example, AWS added new services, including Elastic Cache, and Dynamo DB,   offers several types of EC2 (Elastic Cloud Computing) profiles including  C3 - compute optimized,  R3 - memory optimized; each instance type provides different sets of computer resources measured by vCPUs (vCPU is a hyper-thread of an Intel Xeon core for M3, C3, R3, HS1, G2, and I2).  This process is expected to continue at a faster pace to accommodate the so-called Big Data applications.

\smallskip

\noindent -3. Cloud over-provisioning demands high initial costs and leads to a low system utilization; this strategy is not economically sustainable \cite{Chang10}. {\it Elasticity} allows cloud users to increase or decrease their resource consumption based on their needs; elasticity is now  based on {\it over-provisioning}, assembling pools of resources far  larger than required to satisfy the average needs.   As a result the average cloud server utilization  is in the 18\% to 30\% range \cite{Barroso07, Barroso13}. Power consumption of clouds based on over-provisioning is  excessive  and has a negative ecological impact \cite{Barroso07, Paya15}.  A 2010 survey \cite{Blackburn10} reports that idle or under utilized servers contribute $11$ million tonnes of unnecessary $CO_{2}$ emissions each year and that the total yearly cost for the idle servers is $\$19$ billion.

\smallskip

\noindent -4. The cloud computing landscape is fragmented. CSPs support different cloud delivery models. and this leads to the vendor lock-in; once becoming familiar and storing her data on one cloud it is very costly for the user to migrate to another CSP.

\smallskip

In a large scale-system tensions between local and global objectives exist. These tensions manifest themselves in questions such as: How to balance the individual cost of autonomous servers with global goals e.g., maximizing the CSP payoff?  How to adapt the price for services to the actual demand? How to find an equilibrium between system reconfiguration and continuous system availability? Moreover,  cooperation must reflect the particular characteristics of the physical organization. Locality is important; indeed, communication across multiple layers of the networking infrastructure is less desirable as the latency increases and the bandwidth decreases. The hypothesis of our research is that self-organization and self-management could address these challenges and provide effective means for cloud resource management. 

Informally, self-organization means synergetic activities of elements when no single element acts as a coordinator and the global patterns of behavior are distributed. Self-management  means that individuals can effectively set their own goals, make decisions on how to achieve those goals, plan and schedule their activities independently, and evaluate the progress towards these goals. Self-management can lead to faster and more accurate resource management decisions. 

Self-management as a result of auctions eliminates the need for a system model and requires only local thus, more accurate information about the state of individual components. This approach has the  potential of optimizing the use of resources and allow Cloud Service Providers (CSPs) to offer services at a lower cost for the consumers \cite{Gutierrez10, Lim09}.  Though  the virtues of self-management have long been recognized, there is, to our knowledge, no cloud computing infrastructure, based on self-organizing principles and self-management. This  is in itself proof of the difficulties to apply these concepts in practice. 

Self-management  has to be coupled with some mechanisms for coalition formation allowing autonomous agents, the servers, to act in concert. Autonomous systems have to cooperate to guarantee QoS by distributing and balancing the workload, replicate services to increase  reliability, and  implement other global system policies. Cooperation means that individual systems have to partially surrender their autonomy.

Self-organization cannot occur instantaneously in an adaptive system. It is critical to give the autonomous cloud platforms interconnected by a hierarchy of networks  the time to form coalitions in response to services requests thus, self-management requires an effective reservation system.
Reservations are ubiquitous for systems offering services to a large customer population, e.g., airline ticketing, chains of hotels, and so on. Existing clouds, e.g., the Amazon Web Services, offer both reservations and spot access, with spot access rates lower than those for reservations.

The solution discussed in this paper involves concepts, policies, and algorithms from several well-established areas of economics and computer science: self-organization and self-management of complex systems; coalition formation and virtual organizations; auction theory and practice;  and system organization and computer architecture.  We discuss related work and our contributions in Section \ref{RelatedWork} and in Section \ref{SystemModel} we describe the system model. Algorithms for the formation of sub-coalitions and for clock-proxy auction are the subjects of Sections  \ref{CoalitionFormation} and \ref{CombinatorialAuctionProtocol}, respectively. The results of a simulation experiment and the conclusions of our work are presented in Sections \ref{ProtocolEvaluation}  and \ref{Conclusions}.

\section{Related Work}
\label{RelatedWork}

The present and future challenges outlined in Section \ref{IntroductionAndMotivation} motivate the search for effective and scalable policies and mechanisms for cloud resource management \cite{Bruneo14,  Chaisiri12,  Li13, Mashayekhy15, Niyato11, Samaan14, Wei10}.  In this section we survey some of the research in this area focused on market mechanisms.

\smallskip

{\bf Coalition formation.} Informally, a {\it coalition} is a group of entities which have agreed to cooperate for achieving a common goal. A {\it virtual organization} involves entities that require a communication infrastructure and dedicated software to support their activities. {\it Coalition formation} is a widely used method for increasing the efficiency of resource utilization and for providing convenient means to access these resources \cite{Muller06}. In recent years, the emergence of large-scale electronic markets,  grid and cloud computing, sensor networks, and robotics have  amplified the interest in coalition formation and virtual organizations  \cite{Lerman00, Li02, Sen00}.  For example, self-organization of sensor networks through bottom-up coalition formation is discussed in \cite{Marinescu10, Sims03}.

Different aspects of resource management in computational grids including load balancing, job-allocation, and scheduling, as well as revenue sharing when agents form coalitions or virtual organizations are discussed in \cite{Carroll10, He05, Khan09, Penmatsa06, Subrata08, Zhang03}. Grid resource allocation is modeled as cooperative games \cite{Khan09} or non-cooperative games \cite{Penmatsa06}. Resource co-allocation is presented in \cite{Zhang03}.

There is little surprise that  the interest in coalition formation migrated in recent years from computational grids to cloud resource management. The vast majority of on-going research in this area  is focused on game-theoretic aspects of coalition formation for cloud federations. A {\it cloud federation} is of a set of CSPs collaborating to provide services to a cloud user community.  

A stochastic linear programing game model for coalition formation is presented in \cite{Niyato11}; the authors analyze the stability of the coalition formation among cloud service providers and  show that resource and revenue sharing are deeply intertwined. An optimal VM provisioning algorithm ensuring profit maximization for CSPs is introduced in \cite{Chaisiri12}. 

A  cloud federation formation described as a hedonic game  and focused on the stability and the fairness of the game is discussed in \cite{Mashayekhy15}.  The profit maximization for each federation is formulated as an integer programming problem (IP) and the game is augmented with a preference relation over the set of federations. The paper assumes that the Virtual Machines (VMs)  contributed by each CSP to a federation  are characterized by several attributes, $a \in \mathcal{A}$ including the number of cores, the amount of memory and of secondary storage. The IP problem for CSP $\mathcal{C}_{i}$ in federation $\mathcal{F}$ is formulated as $ \max \sum_{\mathcal{C}_{i} \in \mathcal{F}} \sum_{j=1}^{n} n_{i,j} (p_{j} - c_{i,j})$ subject to the set of conditions $\sum_{j=1}^{n} q_{j}^{a} n_{i,j} \le A_{i},~\forall a \in \mathcal{A}$ and $\sum_{\mathcal{C}_{i} \in \mathcal{F}} n_{i,j} = r_{j}$
with:  $n_{i,j}$ - the number of VMs of type $j$; $p_{j}$ - the price for a VM running an instance of type $j$;  $c_{i,j}$ - the cost of an instance of type $j$ provided by $\mathcal{C}_{i}$; $q_{j}^{a}$ - the quantity of resource of type $a$ in a VM of type $j$;  $A_{i}$ - the total amount of resource of type $a$ offered by $\mathcal{C}_{i}$; and $r_{j}$ - the number of $VM$s of type $j$ requested. The paper adopts a payoff division based on the Banzhaf value \cite{Mashayekhy15}.

A combinatorial coalition formation problem is described in \cite{Li02}. The paper assumes that a seller has a price schedule for each item. The larger the quantity requested, the lower is the price a buyer has to pay for each item; thus, buyers can take advantage of price discounts by forming coalitions. A similar assumption is adopted by the authors of \cite{Lerman00} who investigate systems where the negotiations among deliberate agents are not feasible due to the scale of the system. The paper proposes a macroscopic model and derives a set of differential equations describing the evolution in time of coalitions with a different number of participants. The results show that even a low rate of leaving away participants allows a coalition to achieve a steady state.

An algorithm to find optimal coalition structures in cooperative games by searching through a lattice like the one in Figure  \ref{CoalitionStructuresFig},  was introduced by \cite{Sandholm99}.  A more refined algorithm is described in \cite{Rahwan09}; in this algorithm the coalition structures are grouped according to the so-called {\it configurations} reflecting the size of the coalitions. 

\smallskip

{\bf Auctions.} Auctions are a widely used mechanism for resource allocation \cite{Clarke71,Groves73}. Among the numerous applications of auctions are: the auctioning of airport take-off and landing slots, spectrum licensing by the Federal Communication Commission (FCC), and industrial procurement. An online auction mechanism for resource allocation in computer clouds is presented in \cite{Zhang03}.

 A {\it combinatorial auction} is one where a buyer requires simultaneous access to  a {\it package} of goods. An auction allows the seller to obtain the maximum feasible profit for the auctioned goods; it is organized by an {\it auctioneer} for every {\it request} of a consumer. A {\it proxy} is an intermediary who collects individual bids from the buyers participating at an auction, computes the total cost of the package from the bids, and communicates this price to the auctioneer. A vast literature including \cite{Ausubel02,Ausubel04,Ausubel06,deVries03} covers multiple aspects of combinatorial auctions including bidding incentives, stability, equilibrium, algorithm testing, and algorithm optimality.

 {\it Package bidding} assumes that a seller offers $\mathcal{N}$ different types of items. A buyer bids for packages of items.  A {\it package} is a vector of integers $\mathcal{Z}=\{z_{1}, z_{2}, \ldots, z_{\mathcal{N}} \}$ which indicates the quantity of each item in the package; the price of items is given by  $\mathcal{M}=\{m_{1}, m_{2}, \ldots, m_{\mathcal{N}} \}$.

Package bidding can be traced back to generalized Vickerey auctions based on  the Vickerey-Clarke-Groves mechanisms \cite{Clarke71, Groves73}. In Vickerey auctions a bidder reports its entire demand schedule. The auctioneer then selects the allocation which maximizes the total value of the package and requires a bidder to pay the lowest bid it would have made to win its portion of the final allocation,  considering all other bids.

In an {\it ascending package auction (APA)} there are $\mathcal{K}$ participants identified by an index, $k=0$ is the seller and $k=1,2, \ldots, \mathcal{K}$ are the buyers \cite{Ausubel02}. Each buyer has a {\it valuation} vector $v_{i} = (\nu_{i}(z), z \in [0, \mathcal{M}])$; $\nu_{k}(z)$ represents the value of package $z$ to the bidder $k$. Some of the rules for this type of auction are: all bids are firm, a bid cannot be reduced or withdrawn; the auctioneer identifies after each round the set of the bids that maximize the total price, the so-called {\it provisional winning bids}. The auction ends when a new round fails to elicit new bids; then the provisional winning bids become the winers of the auction.

In an ascending package auction  a bidder can be deterred from bidding for the package she really desires by the threat that competitors could drive prices up; this would threaten the equilibrium. This problem does not exist in {\it ascending proxy auctions} when each bidder instructs a proxy agent to bid on her behalf \cite{Ausubel02}. The proxy accepts as input the bidder's valuation profile and bids following a ``sincere strategy.'' Nash equilibrium can be reached when the bid increments are negligibly small \cite{Ausubel02}.

In a {\it clock auction}  the auctioneer announces prices and the bidders indicate the quantities they wish to buy at the current price. When the demand for an item increases, so does its price until the there is no excess demand. On the other hand, when the offering exceeds the demand, the price decreases \cite{Ausubel02}. In a clock auction the bidding agents see only aggregate information, the price at a given time, and this eliminates collusive strategies and interactions among bidding agents. The auction is {\it monotonic}, the amounts auctioned decrease continually  and this guarantees that the auction eventually terminates.  When the price of a package can be computed as the sum of products of prices and quantities it is said that auction benefits from {\it linear pricing}.

The {\it clock-proxy-auction} is a hybrid auction based on an iterative process with two phases \cite{Ausubel06}. A {\it clock} phase is followed by a {\it proxy} round. During the proxy round the bidders report the values they have submitted to the proxy which in turn submit bids for the package to the auctioneer. A bidder has a single opportunity to report the quantity and the price to the proxy, bid withdrawals are not allowed, and the bids are mutually exclusive. The auctioneer then selects the winning bids that maximize the seller's profit.

\smallskip

\begin{figure}[!ht]
\begin{center}
\includegraphics[width=9.0cm]{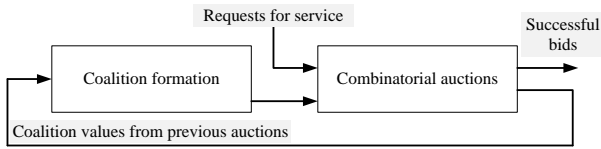}
\end{center}
\caption{A protocol with two stages; feedback about past values of individual coalitions is
used to determine the value of individual coalition structures as shown in Section \ref{CoalitionFormation}.}
\label{FeedbackFig}
\end{figure}

{\bf The contribution of this paper.}  The reservation system we propose has two stages; coalitions of servers are formed periodically during the first and in the second the coalitions participate in  combinatorial auctions organized in each allocation slot. To our knowledge this is the first attempt to address cloud self-organization and resource management based on coalition formation and combinatorial auctions when individual servers learn from past behavior, see Figure \ref{FeedbackFig}.

We discuss coalition formation for a realistic model of the cloud infrastructure, hierarchical organization, while most of the research reported in the literature is focused on coalition formation for cloud federations. The coalition formation problem has different formulations and different constraints in the two cases.  

At this time individual CSPs believe that they have a competitive advantage due to the unique value of their services and are not motivated to disclose relevant information about the inner working of their systems as we have re-discovered when investigating the energy consumption of AWS instances \cite{Paya15}. Thus, the practical realization of cloud federations seems a rather remote possibility \cite{Marinescu13}.  

A rare glimpse at the architecture of a cloud is provided in \cite{Barroso13} and we are taking advantage of it to base our research on a realistic model of the cloud infrastructure. We investigate coalition formation subject to the physical constraints of the hierarchical cloud organization model. As more diverse applications, including Big Data applications, are likely to use computer clouds, the demand for computing resources allocated to a single application will increase and could be considerably larger than any server can provide; only coalitions of servers will be capable to offer such resources.  It is critical for the members of a coalition to  communicate effectively; this requires coalition member  to be in close proximity of each other in a system consisting of a hierarchy of networks with different bandwidth and latency.  This adds additional constrains to the coalition formation protocol.  

To respond to the needs of  increasingly more complex applications consisting of multiple phases and requiring workflow management, CSPs are already offering workflow management services  such as SWS (Simple Workflow Management) and EBS (Elastic Bean Stock) at AWS. Different phases of an application may require coalitions of servers with different types of resources and this is the reason why we decided to investigate combinatorial auctions where packages of items are auctioned.

\section{System Model}
\label{SystemModel}

{\bf System architecture.} We assume a hierarchical organization of the cloud infrastructure similar to the one described in \cite{Barroso13}. A data center consists of multiple warehouse-scale computers (WSCs), each WSC has multiple cells, each cell has multiple racks and each  rack houses multiple servers.   A WSC connects $50,000$ to $100,000$ servers and uses a hierarchy of networks. The servers are housed in racks; typically, the $48$ servers in a rack are connected by a $48$ port Gigabit Ethernet switch. The switch has two to eight up-links which go to higher level switches in the network hierarchy \cite{Barroso13}. The bandwidth to communicate outside the rack is much smaller than the one within the rack; this has important implications for resource management policies and  becomes increasingly difficult to address in systems with a large number of servers.

\smallskip

{\bf Model assumptions.} For simplicity we assume that the racks are homogeneous, they have identical processors with the same number of cores and an identical configuration of GPUs, FPGAs, workflow engines, or  other hardware, the same amount of main storage, cache, and secondary storage.  We also assume that all servers in a rack are identically configured and support the same type of services. The same service may be offered by multiple racks; for example,  multiple racks could offer configurations with GPUs. 

The system we envision supports a {\it reservation system} and {\it spot} resource allocation.  The reservation system has two stages: (A) coalition formation, and (B) combinatorial auctions.
The spot allocation is done through a bidding process for each type of service. The time is quantified, reservations are made  as a result of auctions carried out at the beginning of  each {\it allocation slot} of duration $\tau$; for example, a allocation slot could be one hour.  

\smallskip

{\bf Coalition formation.}  The rational for coalition formation is that applications may need resources  beyond those provided by an individual server. For example,  a Map-Reduce application may require a set of 20 servers during the Map phase to process a data set of several PB (Petabytes). If the algorithm requires the servers to communicate during this phase then the application should start at the same time on all servers and run at the same pace, a condition known as {\it co-scheduling.}  Co-scheduling is only feasible if the set of 20 servers  form a coalition dedicated to the application; moreover, the hardware configuration of the coalition members should be optimal for the algorithms used by the application, e.g., have attached GPUs. 

\smallskip

{\bf  Combinatorial auctions.} Combinatorial auctions allow cloud users making the reservations to acquire packages consisting of coalitions of servers with different types and amounts of resources. Combinatorial auctions are necessary because different phases of an application may require systems with different configurations or systems supporting different functions.
In our previous example the Reduce phase of the Map-Reduce application may require several servers   with a very large amount of secondary storage.

\section{Coalition Formation}
\label{CoalitionFormation}

First, we discuss the formulation of the coalition formation problem as a cooperative game. Then we introduce the algorithms for determining the optimal coalition structure and for coalition formation in the context of our model.  

{\bf Coalition formation as a cooperative game.} The coalition formation is modeled as a cooperative game where the goal of all agents is to maximize the reward due to the entire set of agents. We consider a set of $N$ servers $\{ s_{1}, s_{2}, \ldots, s_{N} \}$, located in the same rack. 

A {\it coalition} $\mathbb{C}_{i}$ is a non-empty subset of $N$. A {\it coalition structure}  is set of $m$ coalitions $\mathbb{S} = \{ \mathbb{C}_{1},  \mathbb{C}_{2}, \ldots, \mathbb{C}_{m} \}$ satisfying the following conditions

\begin{equation}   
\bigcup_{i=1}^{m} \mid \mathbb{C}_{i} \mid = N ~~\text{and}~~ i \ne j~  \Rightarrow ~ \mathbb{C}_{i} \bigcap \mathbb{C}_{j} = \emptyset. 
\end{equation}

\begin{figure*}[!ht]
\begin{center}
\includegraphics[width=14.6cm]{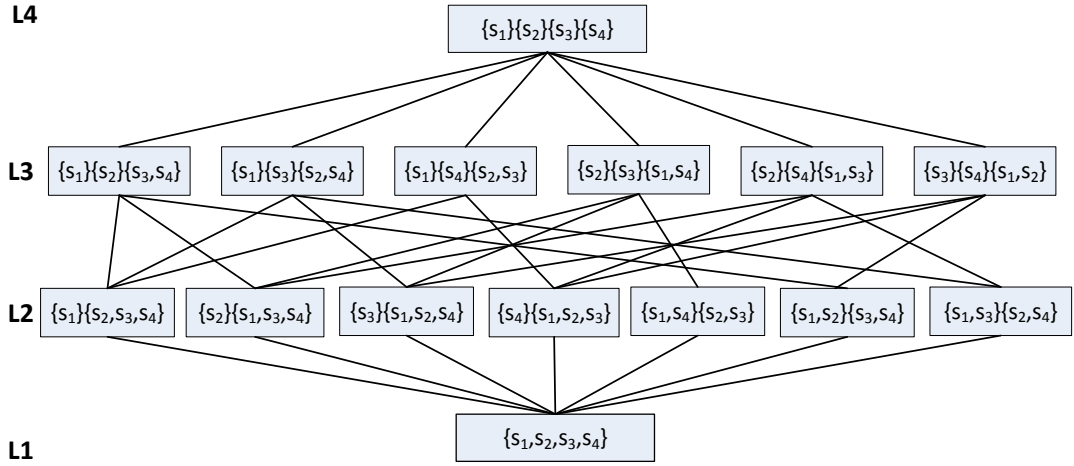}
\end{center}
\caption{A lattice with four levels $L1, L2, L3$ and $L4$ shows the coalition structures for a set of 4 servers, $s_{1}, s_{2}, s_{3}$ and $s_{4}$. The number of coalitions in a coalition structure at level $L_{k}$ is  equal to $k$.}
\label{CoalitionStructuresFig}
\end{figure*}

Figure \ref{CoalitionStructuresFig} shows a lattice representation of the coalition structures for a set of four servers $s_{1}, s_{2}, s_{3}$ and $s_{4}$. This lattice has four levels, $L1, L2, L3$ and $L4$  containing the coalition structures with $1, 2, 3$ and $4$ coalitions, respectively. In general, the level $k$ of a lattice contains all coalition structures with $k$ coalitions; the number of of coalitions structures at level $k$ for a population of $N$ agents is given by the Sterling Number of Second Kind:

\begin{equation}
\mathcal{S}(N,k) = { 1 \over k! } \sum_{i=0}^{k} (-1)^{i} {k \choose i} (k-i)^{N}.
\end{equation}
In the case illustrated in Figure \ref{CoalitionStructuresFig} $N=4$ and the number of coalition structures at levels $L1 - L4$ are $1, 7, 6, 1$, respectively.\footnote{For $N=5$ and $N=6$ the Stirling Numbers of the Second Kind are respectively $1, 15, 25,  10, 1 $ and $1, 31, 90, 65, 15,1$.} The total number of coalition structures with $N$ agents is called the Bell number 

\begin{equation}
\mathcal{B}(N) = \sum_{k=0}^{N} \mathcal{S}(N,k) = \sum_{k=0}^{N} { 1 \over k! } \sum_{i=0}^{k} (-1)^{i} {k \choose i} (k-i)^{N}.
\end{equation}
The number of coalitions structures increases exponentially with the number of agents. For example, for $N=40$, a typical number of servers in a rack, the logarithm of the number of coalition structures is close to $10^{35}$ and $\mathcal{S}(40,14)=3.5859872255621803491428554E+34$. The logarithm of number of coalitions is close to $E+{10}$.  

Searching for the optimal coalition structure $\mathbb{C}$ is computationally challenging due to the size of the search space. The first step for determining the optimal coalition structure is to assign a  {\it value} $v$ reflecting the utility of each coalition.  The second step is the actual coalition formation.

{\bf Rack-level coalition formation.} Recall from Section \ref{SystemModel} that in our model a rack is homogeneous, all servers have an identical configuration. This realistic assumption simplifies considerably the complexity of the search for an optimal coalition structure as the servers are indistinguishable from one another. 

The second important observation is that we have a system with two stages and feedback, see Figure \ref{FeedbackFig}. In the second stage the coalitions created during the first stage are included in successfully auctioned packages thus, we can determine precisely the value of all coalitions structures. The third  important observation is that only {\it available servers},  servers with no commitments for the current slot, can participate to coalition formations and then to the auction organized in that slot; call $N_{a} \le N$ the number of available servers.

An elected {\it rack leader}  collects information about all {\it successful coalitions} - coalitions that have been included in packages auctioned successfully during a window of $w$ successive past allocation slots. The current rack-leader records an entry for the corresponding {\it partial coalition structure (PCS)} including $n_{k}$ - the {\it coalition size}, $m_{k}$ - the {\it multiplicity} of occurrence,  the value  $\bar{v}_{k}$ calculated as the average price over all auctions when a PCS including a coalition of size $n_{k}$ was part of a {\it package} successfully auctioned during the past $w$ allocation slots.

Call $\mathcal{L}$ the {\it PCL-list}. For a window of size $w$ the list $\mathcal{L}$ is the list of all triplets $\mathcal{L}_{k} = [n_{k}, m_{k}, \bar{v}_{k}]$ ordered first by $1 \le n_{k} \le N_{a} $ then by $m_{k}$.  The list includes only entries $\mathcal{L}_{k}$ with $\bar{v}_{k} > 0$. Given $N_{a}$ a {\it coalition structure (CS)} $\mathbb{S}_{k}$  among the entries $\mathcal{L}_{k1}, \mathcal{L}_{k2},... \ldots, \mathcal{L}_{kn}$ is {\it feasible} if $\sum_{j} n_{k} \times m_{k} = N_{a}$.  Then the value of the coalition structure $\mathbb{S}_{k}$ is $v_{k} = \sum_{j} \bar{v}_{j}$. Note that we force the formation of coalitions involving all available servers. An example of a PCS list $\mathcal{L}$ follows
\begin{verbatim}
----------------------------------------------
a  [1,4,35]    \* 4 PCS of 1-server {s} 
b  [1,15,682]  \ *15 PCS of 1-server {s} 
       .........
c  [2,3,78]     \* 3 PCS of  2-servers {s,s}
       ........
d  [3,2,502]   \* 2 PCS of 3-servers {s,s,s} 
e  [3,4,812]   \* 4 PCS of 3-servers {s,s,s} 
       .........   
f  [16,1,751]  \* 1 PCS of 16-servers {s,...s}
g  [16,2,740]  \* 2 PCS of 16-servers {s,...s}
       .........
-----------------------------------------------
\end{verbatim}
In this example some of the feasible coalitions structures when $N_{a} = 16$ are: $\mathbb{S}_{g} $ with $v_{g}=751$; $\mathbb{S}_{a,b}$ with  $ v_{a,b} = 35+682=712 $; $\mathbb{S}_{a,e}$ with $v_{a,e} = 35+ 812=837$; $\mathbb{S}_{a,c,d}$ with $ v_{a,c,d} = 35 +78 + 502 = 615 $, and so on. Note that the value of  a coalition reflects also the length of time the coalition was active in response to successful auction. We see that a PCS of $15$ coalitions of 1 server have been active  for larger number of slots than a PCS of $4$ coalitions of 1 server. The value attributed to a coalition of $k$ servers is a distributed equally among the servers; the value of a package of several coalitions auctioned successfully is divided among the coalitions based on the resource supplied by each one of them.

\smallskip

{\bf Coalition formation.}  The protocol for coalition formation proceeds as follows:

\begin{enumerate}
\item
Server $s_{i}$  sends to the current rack leader:
\begin{enumerate}
\item
A vector $([\nu_{i}^{1},\beta_{i}^{1}], [\nu_{i}^{2},\beta_{i}^{2}],\ldots ....[\nu_{i}^{N}, \beta_{i}^{N}])$ with $\nu_{i}^{k}, 1 \le k \le N$ the total value due to the participation of $s_{i}$ in successful coalitions, of $k$ servers and $\beta_{i}^{k}$ a bit vector with $w$ components with $\beta_{i}^{k,j}=1$ if $s_{i}$ was included in a successful coalition of $k$ servers in slot $j$ of window $w$. 
\item
Availability, $a_{i}=1$ if available, $0$ otherwise.
\end{enumerate}

\item
After receiving the information from all servers the current rack leader:

\begin{enumerate}
\item
Determines $N_{a} = \sum_{i=1}^{N} a_{i}$.
\item
Computes $m_{k} = \sum_{i=1}^{N_{a}} \sum_{j=1}^{w} \beta_{i}^{k,j}, 1 \le k \le N$.
\item
Computes $\bar{v}_{k} = \sum \nu_{i}^{k}$
\item
Computes the optimal coalition structure.
\item
Assigns a server to coalition of size $k$ a  based on the values  $\nu_{i}^{k}$.
\item
Chooses the best performer as the next coalition leader. The best performer is the one with the largest 
value $\sum_{j} \nu_{i}^{j}$.
\end{enumerate}
\end{enumerate}

Finding the optimal CS requires at most $L$ operations with $L$ the size of the PCL-list.  The system starts with a predetermined coalition structure and coalition values.

\section{A Reservation System Based on a Combinatorial Auction Protocol}
\label{CombinatorialAuctionProtocol}

The protocol introduced in  this section targets primarily the {\it IaaS} cloud delivery model represented by Amazon Web Services (AWS). Reservation systems are currently used by CSPs. For example, AWS supports reservations as well as spot allocation and offers a limited number of instance families, including M3 (general purpose), C3 (compute optimized), R3 (memory optimized), I2 (storage optimized), G2 (GPU) and so on. An instance is a package of system resources; for example, the {\tt c3.8xlarge} instance provides 32 vCPU, 60 GiB of memory, and $2 \times 320$ GB of SSD storage. The  resources auctioned are supplied by coalitions of servers in different racks and the cloud users request packages of resources.

The combinatorial auction protocol is inspired by the clock-proxy auction \cite{Ausubel06}. The clock-proxy auction has a clock phase, where the price discovery takes place, and a proxy phase, when bids for packages are entertained. In the original clock-proxy auction there is one seller and multiple buyers who bid for packages of goods. 

For example, the airways spectrum in the US is auctioned  by the FCC and communication companies bid for licenses.  A package consist of multiple licenses; the quantities in these auctions are the bandwidth allocated  times the population covered by the license. Individual bidders choose to bid for packages during the proxy phase and pay the prices they committed to during the clock phase.

Our protocol supports auctioning service packages; a packages consist of combinations of services in one or more time slots. The items sold are services advertised by coalitions of autonomous servers and the bidders are the cloud users. Each service is characterized by
\begin{enumerate}
\item
A {\it type} describing the resources offered and the conditions for service,
\item
The time slots when the service is available.
\end{enumerate}

\begin{figure*}[!ht]
\begin{center}
\includegraphics[width=12.0cm]{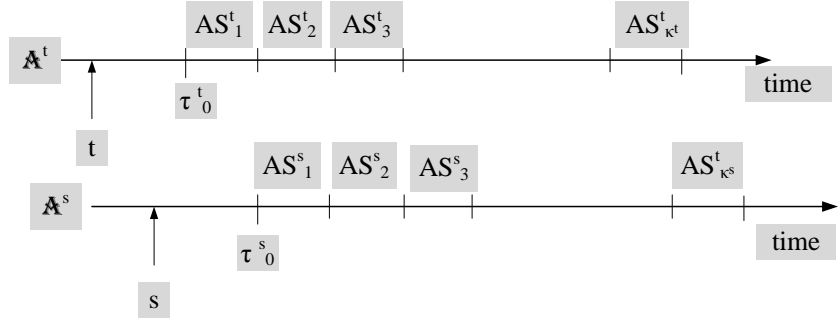}
\end{center}
\caption{Auctions $\mathbb{A}^{t}$ and $\mathbb{A}^{s}$ conducted at times $t$ and $s$, respectively. $\tau^{t}_{0}$ and $\tau^{s}_{0}$ are the start of the first allocation slots, $AS^{t}_{1}$ and $AS^{s}_{1}$ of the two auctions. The number of slots auctioned in each case are $\kappa^{t}$ and $\kappa^{s}$, respectively.}
\label{AuctionTimingFig}
\end{figure*}

{\bf Protocol specification.}  The terms used to describe the protocol are discussed next. An {\it allocation slot} (AS) is a period of fixed duration, e.g., one hour, that can be auctioned. An {\it auction}, $\mathbb{A}^{t}$, is organized at time $t$ if there are pending reservation requests which require immediate attention. Figure \ref{AuctionTimingFig} shows two consecutive auctions at times $t$ and $s$; during the first slot of auction $\mathbb{A}^{t}$ new reservation requests are received and the allocation slot $AS^{t}_{2}$ is not fully covered; this slot becomes $AS^{s}_{1}$ for $\mathbb{A}^{s}$.

A service $\mathcal{A}$ is described by a relatively small number of {\it attributes}, $\{a_{1}, a_{2}, \ldots,  \}$. Each attribute $a_{i}$ can take a number of distinct values, $v_{i} =\{v_{i,1},v_{i,2}, \ldots  \}$. The first attribute is the coalition size or equivalently the number of vCPS provided; other attributes could be the type of service and server architecture with two values ``32-bit'' and ``64-bit;'' another attribute could be ``organization'' with values ``vN'' (von Neumann), ``DF'' (data-flow), or ``vN-GPU'' (vN with graphics co-processor).

Call $\mathcal{S}^{t}$ the set of services the clients want to reserve during auction $\mathbb{A}^{t}$

\begin{equation}
\mathcal{S}^{t} =\{ S^{t}_{1}, S^{t}_{2}, \ldots , S^{t}_{\nu^{t}} \}~~\text{with}~~S^{t}_{i}=[sId, (a_{j},v_{j,k})]
\end{equation}
A {\it reservation  bundle,}  $\alpha^{t}_{i,j} \subset \mathcal{S}^{t}$, is the set of services requested by client $i$ in slot $j$ of auction $\mathbb{A}^{t}$

\begin{equation}
\alpha^{t}_{i,j}=\{ (S^{t}_{i,j,1}, r^{t}_{i,j,1}), (S^{t}_{i,j,2}, r^{t}_{i,j,2}) , \ldots \}
\end{equation}
with $r^{t}_{i,j,l}$ a measure of the quantity; for example, if the attribute is ``service intensity'' the quantity is the number of vCPUs.

An {\it advertised  bundle,}  $\beta^{t}_{k,j} \subset \mathcal{S}^{t}$, is the set of services advertised by coalition $k$ in slot $j$ of auction $\mathbb{A}^{t}$
\begin{equation}
\beta^{t}_{k,j}=\{ (S^{t}_{k,j,1}, q^{t}_{k,j,1}, p^{t}_{k,1}),(S^{t}_{k,j,2}, q^{t}_{k,j,2}, p_{k,2})  \ldots \}
\end{equation}
with $q^{t}_{k,j,l}$ a measure of the quantity of service $l$ and $p_{k,l}$  the price per ECU of service $S^{t}_{l}$ determined by coalition $k$. A {\it package}, $\mathcal{P}^{t}_{i}$ is a set of reservations for services requested by client $i$ for slots $j_{1}, j_{2}, \ldots$ during auction $\mathbb{A}^{t}$.

\begin{equation}
\mathcal{P}^{t}_{i} =\{\alpha^{t}_{i,j_{1}}, \alpha^{t}_{i,j_{2}}, \ldots  \}
\end{equation}

\begin{figure}[!ht]
\begin{center}
\includegraphics[width=8.5cm]{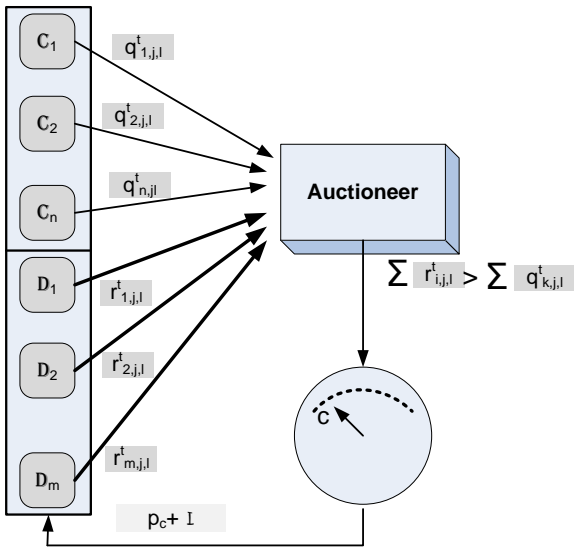}
\end{center}
\caption{The clock phase for service $S^{t}_{l}$ and slot $j$. The starting price is $p_{l}^{0}$ given by Equation \ref{startingPrice}. The clock advances and the price increases from $p_{c}$ to $p_{c} + \mathcal{I}$ when the available capacity at that price given by Equation \ref{availableCapacity} is exhausted; the demand is given by Equation \ref{availableCapacity}.}
\label{ClockAlgorithmFig}
\end{figure}

{\bf The clock phase.} Figure \ref{ClockAlgorithmFig} illustrates the basic idea of a clock phase: the auctioneer announces prices and the bidders indicate the quantities they wish to buy at the current price. When the demand for an item increases, so does its price until there is no excess demand; on the other hand, when the offering exceeds the demand, the price decreases.

During the clock phase of auction $\mathbb{A}^{t}$ the price discovery is done for each time slot and for each type of service; a clock runs for each one of the $\kappa^{t}$ slots and for each one of the  $\nu^{t}$ services.  Next we describe the clock phase for service $S^{t}_{l}$ in slot $j$. Assume that there are $n$ coalitions $\mathbb{C} =\{ \mathbb{C}_{1}, \mathbb{C}_{2}, \ldots, \mathbb{C}_{n} \}$ offering the service and $m$ requests for reservations from clients  $\mathbb{D} =\{ \mathbb{D}_{1}, \mathbb{D}_{2}, \ldots, \mathbb{D}_{m} \}$.

A clock auction starts at clock time $t=0$ and at price per unit of service for $S_{l}$

\begin{equation}
\label{startingPrice}
p_{l}^{0} = \min_{\mathcal{C}_{k}} \{p_{k,l}\}
\end{equation}
Call $\mathcal{C}_{0}$ the available capacity at this price and $\mathcal{D}_{0}$ the demand for service $S^{t}_{l}$ offered at price $p_{l}^{0}$ in slot $j$

\begin{equation}
\label{availableCapacity}
\mathcal{C}_{0} = \sum_{k=1}^{n} q^{t}_{k,j,l}
~~\text{and}~~
\mathcal{D}_{0} = \sum_{i=1}^{m} r^{t}_{i,j,l}.
\end{equation}
If $\mathcal{C}_{0} < \mathcal{D}_{0}$ the clock $c$ advances and the next price per unit of service is

\begin{equation}
p_{l}^{1} = p_{l}^{0} + \mathcal{I}
\end{equation}
with $\mathcal{I}$ the price increment decided at the beginning of auction. There is an ample discussion in the literature regarding the size of the price increment; if too small, the duration of the clock phase increases, if too large, it introduces incentives for gaming \cite{Ausubel06}.

The process is repeated at the next clock value starting with the new price. The clock phase for service $S^{t}_{l}$ and slot $j$ terminates when there is no more demand.

\begin{figure*}[!ht]
\begin{center}
\includegraphics[width=16cm]{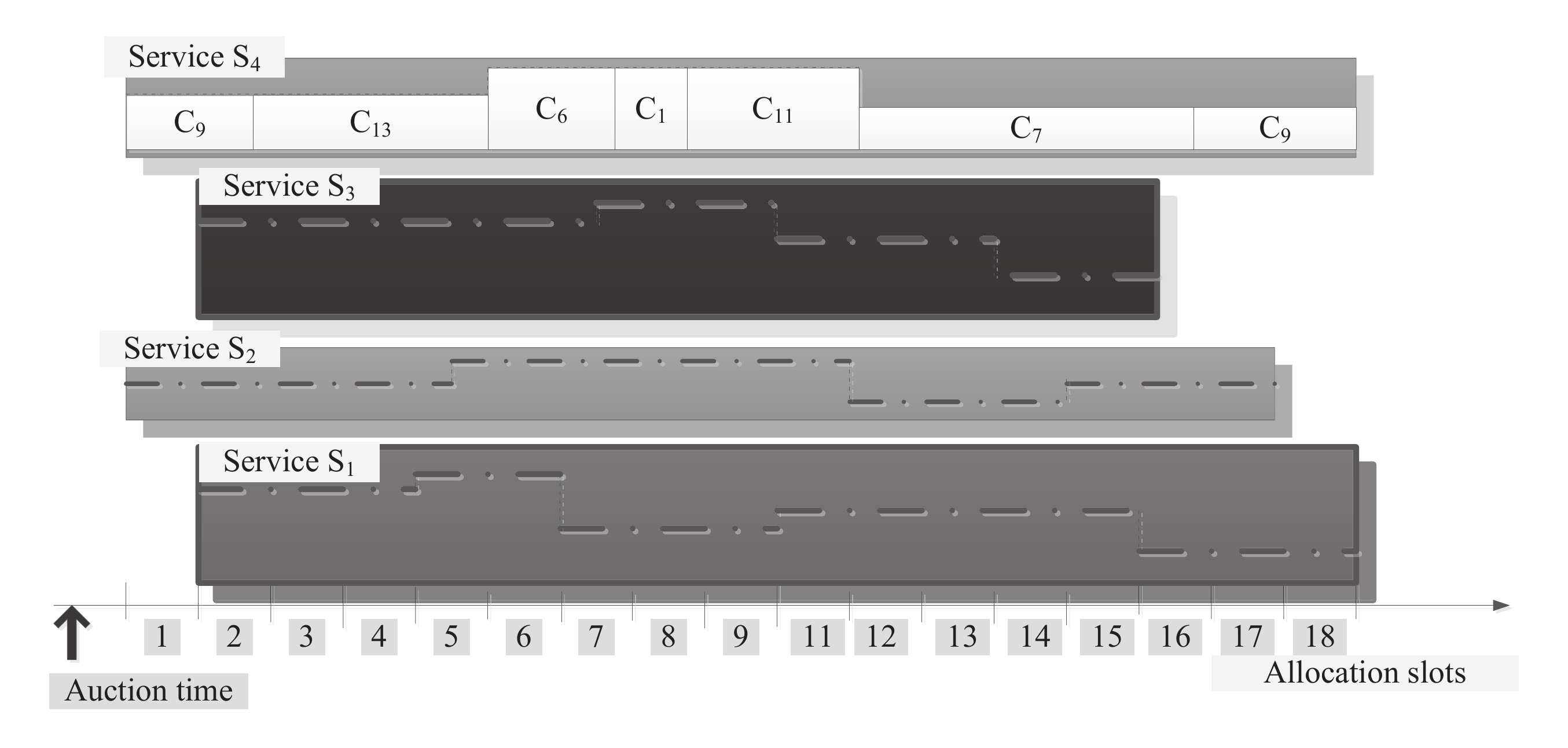}
\end{center}
\caption{A snapshot at the end of the preliminary rounds of the proxy phase when there are four services offered and the auction covers $18$ allocation slots. Dotted lines represent the quantity of service with provisional winners. Only the provisional winers for $S_{4}$ are shown, the clients labeled as $C_{9}, C_{13}, C_{6}, C_{1}, C_{11}, C_{7}$ and $C_{9}$.}
\label{ServicesAndSlotsFig}
\end{figure*}

{\bf The proxy phase.} In a traditional clock-proxy auction the bidders do not bid directly, they report to a proxy the price and the quantity of each item in the package they desire. The proxy then bids in an ascending package auction.

In our application, the proxy phase of the auction consists of multiple rounds. The auction favors bids for long runs of consecutive slots when the service is provided by the same coalition. This strategy is designed to exploit temporal and spatial locality.

The auction starts with the longest runs and the lowest price per slot and proceeds with increasingly shorter  runs and diminished incentives. Once a run of consecutive slots is the subject of a provisional winning bid, all shorter runs of slots for that particular service are removed from the  coalition offerings.

During the first round only the longest run of consecutive slots for each one of the services offered by the participating coalitions is auctioned and only bidders that have committed to any of the slots of the run are allowed to bid. The price per slot for the entire run is the lowest price for any slot of the run the bidder has committed to during the clock phase of the auction. If there are multiple bids for service $S^{t}_{l}$ the {\it provisional winner} is the one providing the largest revenue for the coalition offering the service.

If $\kappa^{t}_{l}$ is the longest run of consecutive slots for service $S^{t}_{l}$ auctioned in the first round then, in the second round, a shorter run of $\kappa^{t}_{l}-1$ slots is auctioned. The price for the entire run equals the second lowest price for any slot of the run the bidder has committed to during the clock phase of the auction times the number of the time slots in the run.

The length of the consecutive slot runs auctioned decreases and the incentives diminish after each round. The preliminary rounds end with the auction of a single slot for each service. At the end of the preliminary round each bidder is required to offer the price for the slot committed to during the clock phase. Figure \ref{ServicesAndSlotsFig} depicts a plausible snapshot at the end of the preliminary rounds of the proxy phase when four services $\mathcal{S}_{1}, \mathcal{S}_{2}, \mathcal{S}_{3}$ and $\mathcal{S}_{4}$, are offered  and shows the provisional winners for service $\mathcal{S}_{4}$.

During the final round the bidders reveal the packages they want to reserve; these packages include only the provisional winners from the preliminary slots. Once all provisional winning bids for services in a reservation request are known, the auctioneer chooses the package that best matches the consumer's needs and, at the same time maximizes the profit for the cloud service provider. The {\it coalition} for a reservation request consists of the set of coalitions that provide the services in the winning package.

In this auction all bids are firm, they cannot be withdrawn. The auction is monotonic, the length of runs of consecutive slots auctioned decreases continually; this guarantees that the auction eventually terminates. Linear pricing guarantees that the price of any package can be computed with ease.

\smallskip

{\bf The effectiveness of the protocol} is captured by several metrics including:

\smallskip

\noindent -1. The {\it customer satisfaction index} - percentage of reservation requests fully or partially satisfied in each allocation slot given the total number of requests.

\smallskip

\noindent -2. The {\it service mismatch index} - percentage of services requested but not offered in each allocation slot given the total number of services in that slot.

\smallskip

\noindent -3. The {\it service success index} - percentage of services used in each allocation slot given all services offered in that slot.

\smallskip

\noindent -4. The {\it capacity allocation index} - percentage of the capacity offered but not auctioned in each allocation slot given the capacity offered in that slot.

\smallskip

\noindent -5. The {\it overbidding factor} - percentage of slots with a provisional winner that have not been included in any package given all slots offered at the beginning of the auction.

\smallskip

\noindent -6. The {\it temporal fragmentation index} - percentage  of services successfully auctioned in non-consecutive slots given all services successfully auctioned.

\smallskip

\noindent -7. The {\it additional profit index} - percentage of additional profit of coalitions involved in the auction (the difference of the actual price obtained at the auction and the price demanded by the coalition) relative to the price demanded by the coalition.

\smallskip

{\bf Limitations and vulnerabilities.} The protocol is fairly complex and has at least one vulnerability. A bidder may be the provisional winner of services in slots not included in its winning package; such services will remain unassigned during the current auction. A solution is to penalize{\it excess bidding activity} and charge the bidder a percentage of the costs for these services. Another alternative  is to include, in a reservation request, a set of ``substitute services'' for a service $S_{i}$. Then, during the last round of the proxy phase, the auctioneer could try to match services having provisional winners with unsatisfied requests for services.

The capacity offered, but not auctioned in each slot is available for {\it spot allocation} thus, it
has the potential to be used, rather then being wasted. The capacity of a coalition left uncommitted  at the end of the auction $\mathbb{A}^{t}$ for $AS^{t}_{1}$, the first slot of the auction, is then available for {\it spot allocation} at a price equal to $p_{k,l}$, while the free capacity in slots starting with $AS^{t}_{2}$ can be offered at the next auction if this auction takes place before the beginning of the slot. This capacity is measured by the {\it spot allocation opportunity index.}

\section{Protocol Analysis and Evaluation}
\label{ProtocolEvaluation}

We report on the results of our simulation experiments to gain some insight into the proxy phase of the clock-proxy auction of the $PC^{2}P$ protocol. The system we wish to evaluate requires the description of the environment in which the auction takes place, the reservation requests, and the services offered:

\begin{enumerate}
\item
The environment elements: $n$ - the number of coalitions offering services in this round;
$m$ - the number of clients; and $\kappa$ - the number of slots auctioned.
\item
The package $j$ requested by client $i$: $\alpha_{i}^{n}$ - the number of services in the package; the slots desired by the service $S_{k}$, ordered by the length of the run of consecutive slots; $r_{k,j}$ - the intensity of service $S_{k}$ in slot $j$;  $p_{i,j}$ - the price per unit of service for slot $j$ if client $i$ was a provisional winner of that slot during the clock phase.
\item
The service $S_{k}$ provided by coalition $\mathbb{C}_{k}$ includes: $\gamma_{k}$ - the largest run of consecutive slots for each offered service $S_{k}$; the profile of the service $S_{k}$ - the slots offered ordered by the length of consecutive slots, when it is available; $q_{k,j}$ - the quantity of service $S_{k}$ offered in slot j; and $p_{k}$ - the price per unit of service offered by coalition $\mathbb{C}_{k}$.
\end{enumerate}

For simplicity, we assume that a coalition offers one service only and the number of services is  $\nu < n$. We also assume that all platforms have a maximum capacity of $100$ vCPUs and that $q_{k,j}$, the quantity of service $S_{k}$ offered for auction, and $r_{k,j}$, the quantity of $S_{k}$ requested in slot $j$ are the same for all the slots of an offered/requested run. The number of slots auctioned is fixed, $\kappa=50$.

The range and the distribution of parameters for the protocol evaluation are chosen to represent typical cases. The parameters of the simulation are random variables with a uniform distribution:

\smallskip

\noindent  -a. The number of coalitions and  clients requesting reservations, $n$ and $m$, respectively; the interval is $[200-250]$.
\smallskip

\noindent  -b. The number of services offered and requested $\nu$; the interval is $[10-20]$.

\smallskip

\noindent  -c. The number of clients bidding for each service in a given slot; the interval is $[0-4]$.

\smallskip

\noindent  -d. The capacity offered for auction for a service in a given slot; the interval is $[60-90]$ vCPUs.

\smallskip

\noindent  -e. The services offered by a coalition; the interval is $[1 - \nu]$.

\smallskip

\noindent  -f. The number of consecutive slots a service is offered in; the interval is $[1 - \kappa]$.

\smallskip

\noindent  -g. The number of services in the package requested by a client; the interval is $[1 - 3]$.

\smallskip

\noindent  -f. The number of consecutive slots of the services in the package requested by a client; the interval is $[1 - \kappa]$.

\smallskip

We also randomly choose the slots when the client is the provisional winner. The evaluation process consists of the following steps:

\smallskip
\noindent A. \underline{Initialization.}

\smallskip

\noindent B. \underline{Preliminary rounds.} Carry out $\gamma$ preliminary rounds with
$\gamma= \max_{k} \gamma_{k}$.

\begin{itemize}
\item
In the first preliminary round auction auction $\kappa_{1}$ slots of service $S_{1}$,  $\kappa_{2}$ slots of service $S_{2}$, and so on.
\item
Identify the first slot of each run and the reservation request that best matches the offer.
\item
Identify the provisional winners if such  matches exist and remove the corresponding runs from the set of available runs. A match exists if the run consists of the same number of slots or is one slot longer than requested and if the capacity offered is at least the one required by the reservation request. For services without a match, remove the last slot, add both the shorter run and the last slot to the list of available runs.
\item
Continue this process until only single slots are available.
\end{itemize}

\noindent C. \underline{Final round.} In this round we:
\begin{itemize}
\item
Identify the packages for each client and if multiple packages exist determine the one which best matches the request.
\item
Compute the cost for the winning package for each client.
\end{itemize}

\begin{figure*}[!ht]
\begin{center}
\includegraphics[width=8.7cm]{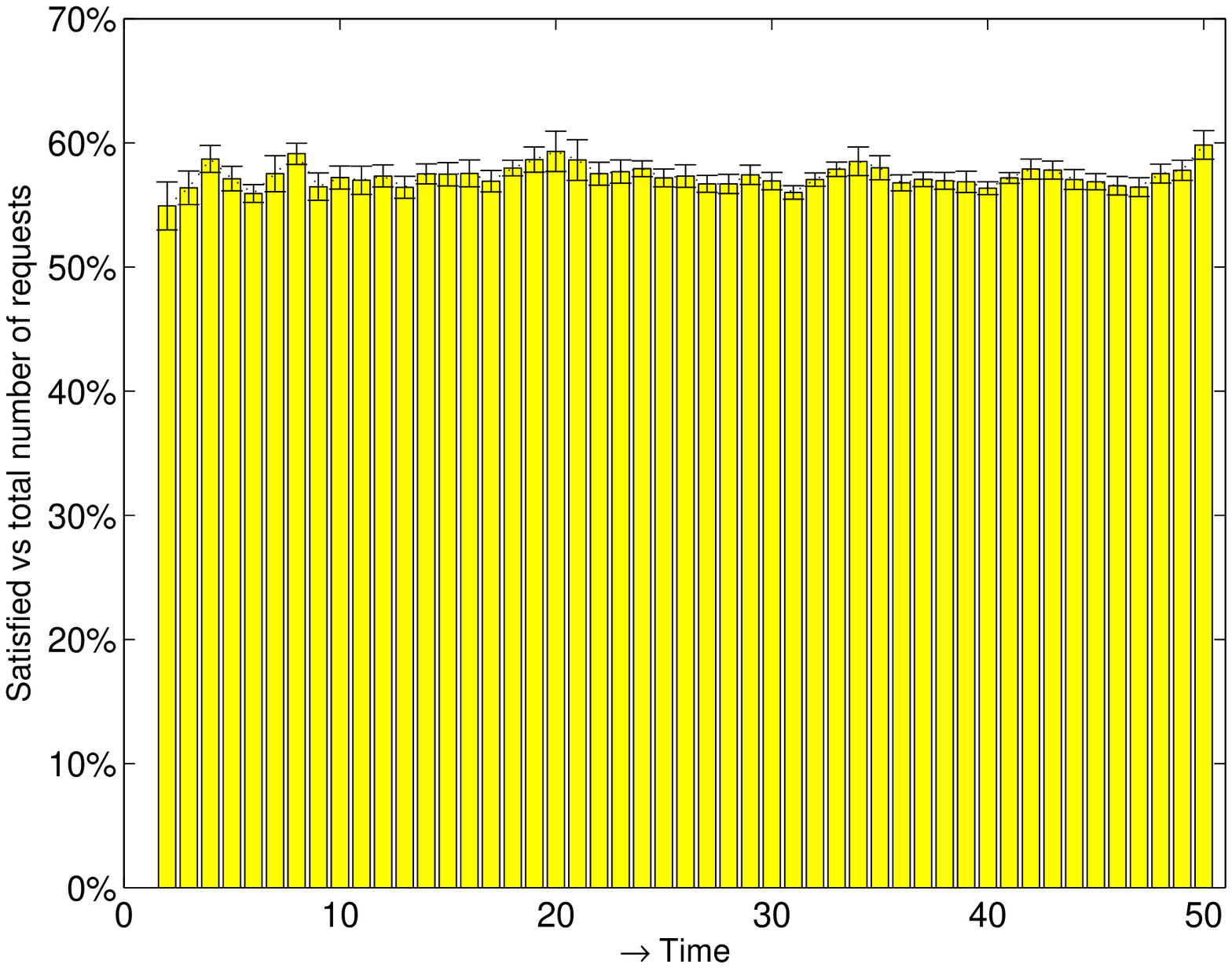}
\includegraphics[width=8.7cm]{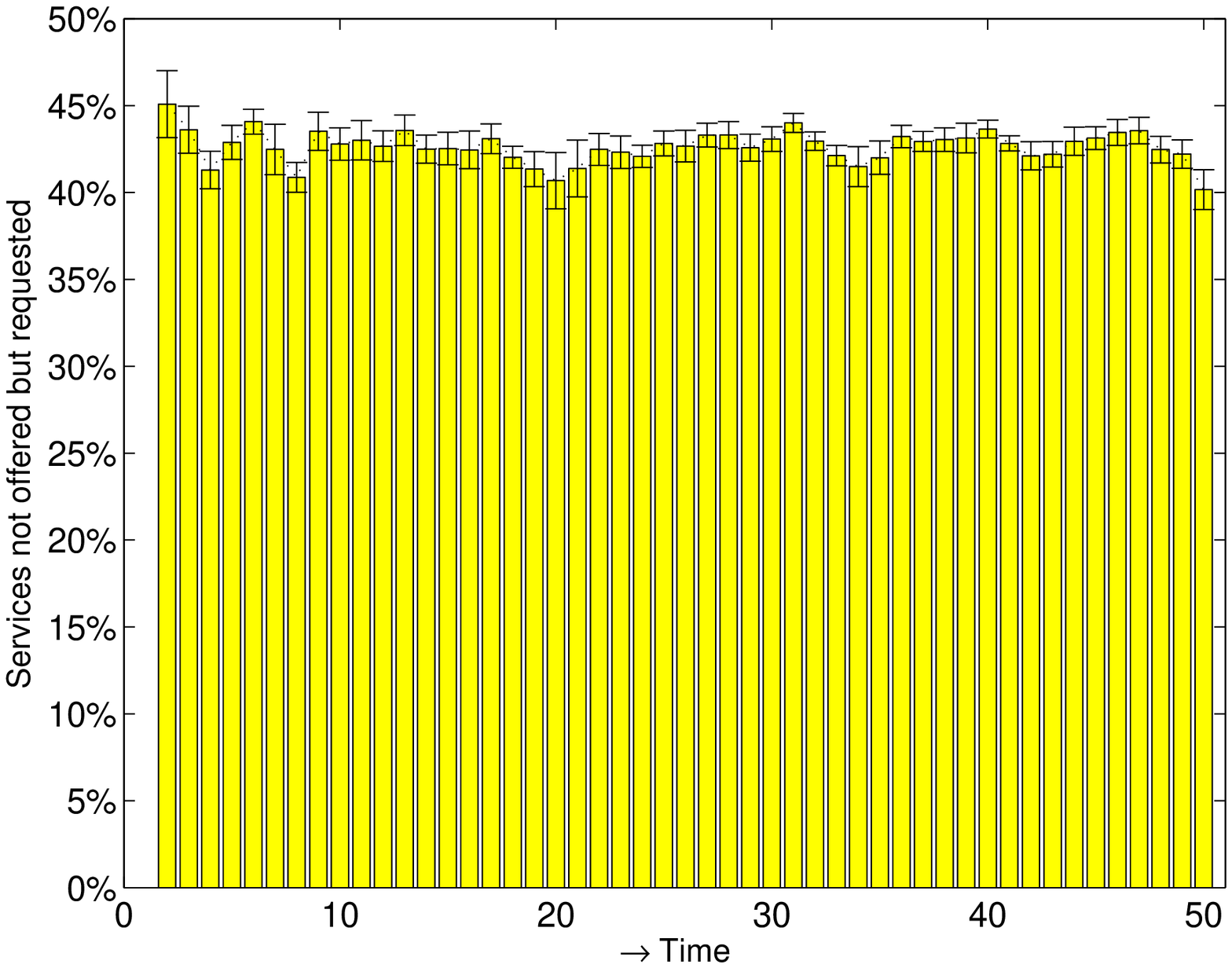}\\
(a)~~~~~~~~~~~~~~~~~~~~~~~~~~~~~~~~~~~~~~~~~~~~~~~~~~~~~~~~~~~~~(b)\\
\includegraphics[width=8.7cm]{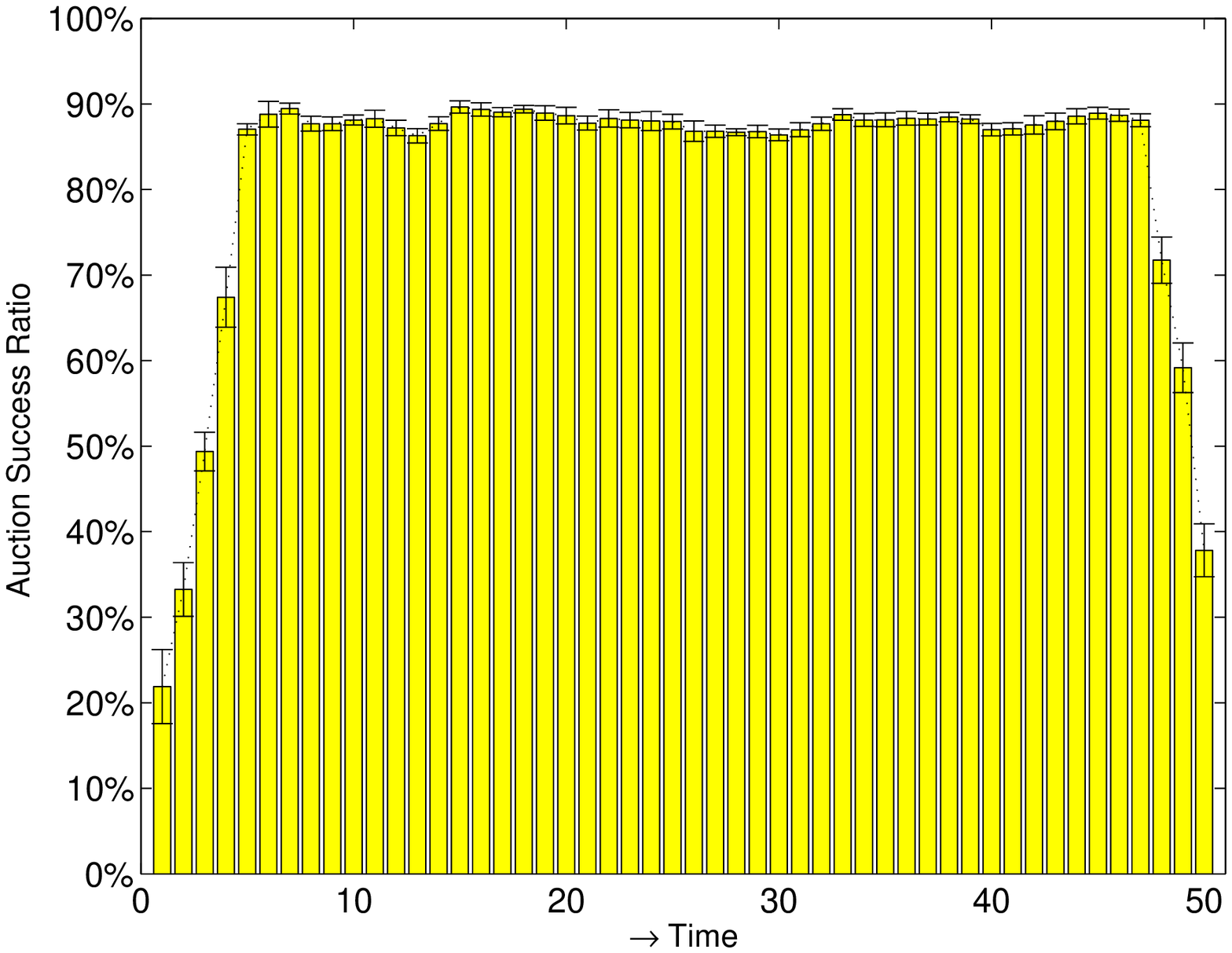}
\includegraphics[width=8.7cm]{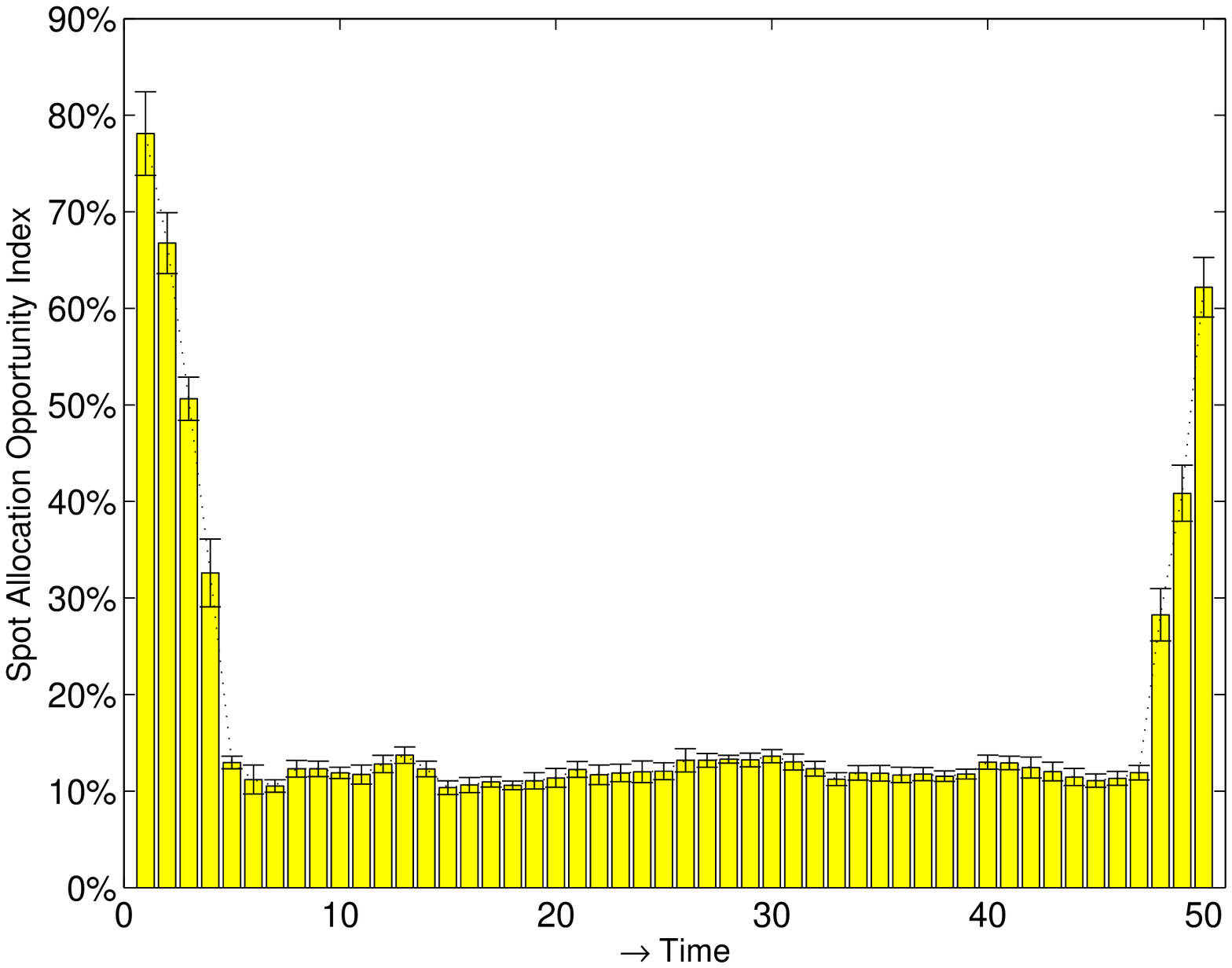}\\
(c)~~~~~~~~~~~~~~~~~~~~~~~~~~~~~~~~~~~~~~~~~~~~~~~~~~~~~~~~~~~~~(d)\\
\includegraphics[width=8.7cm]{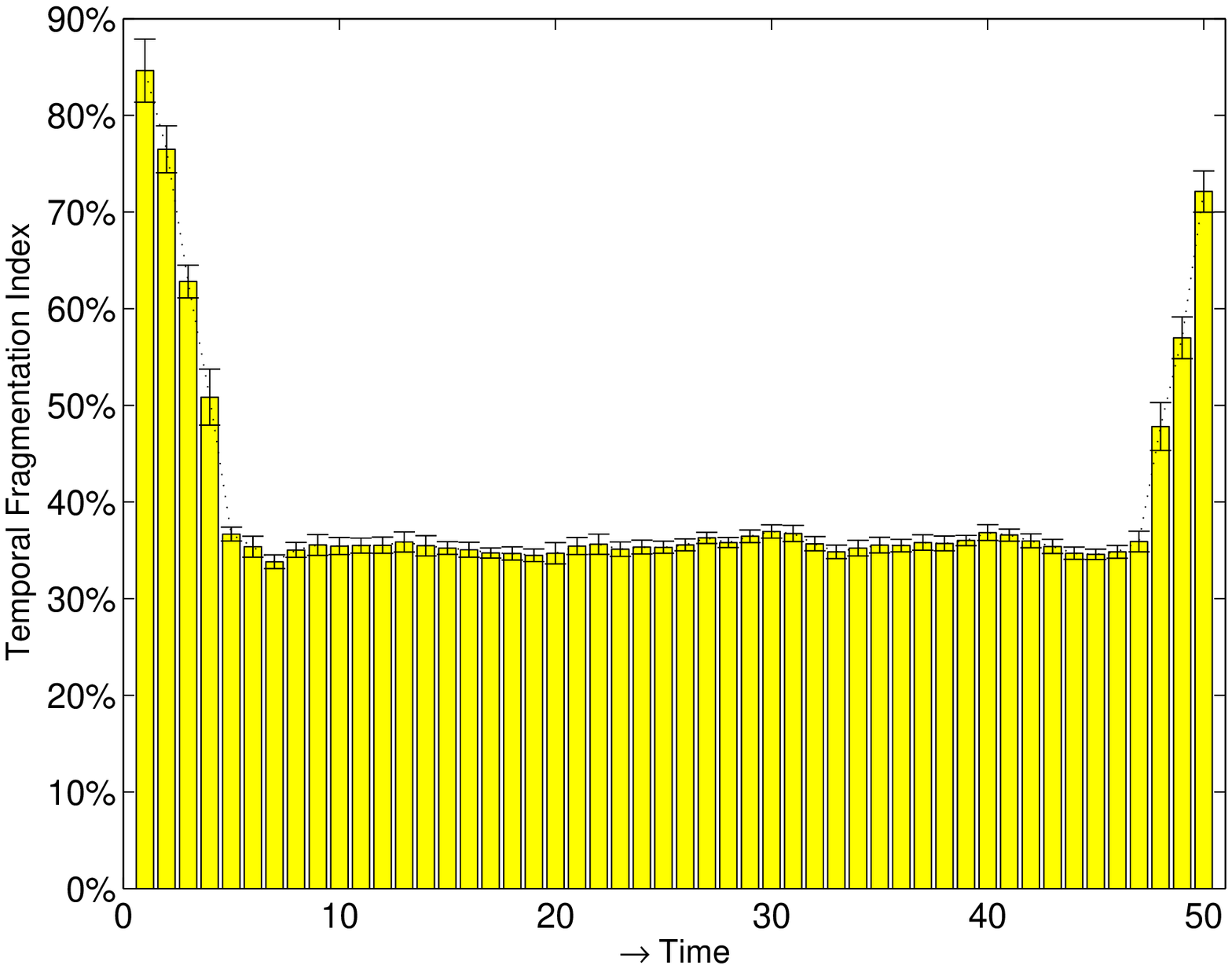}
\includegraphics[width=8.7cm]{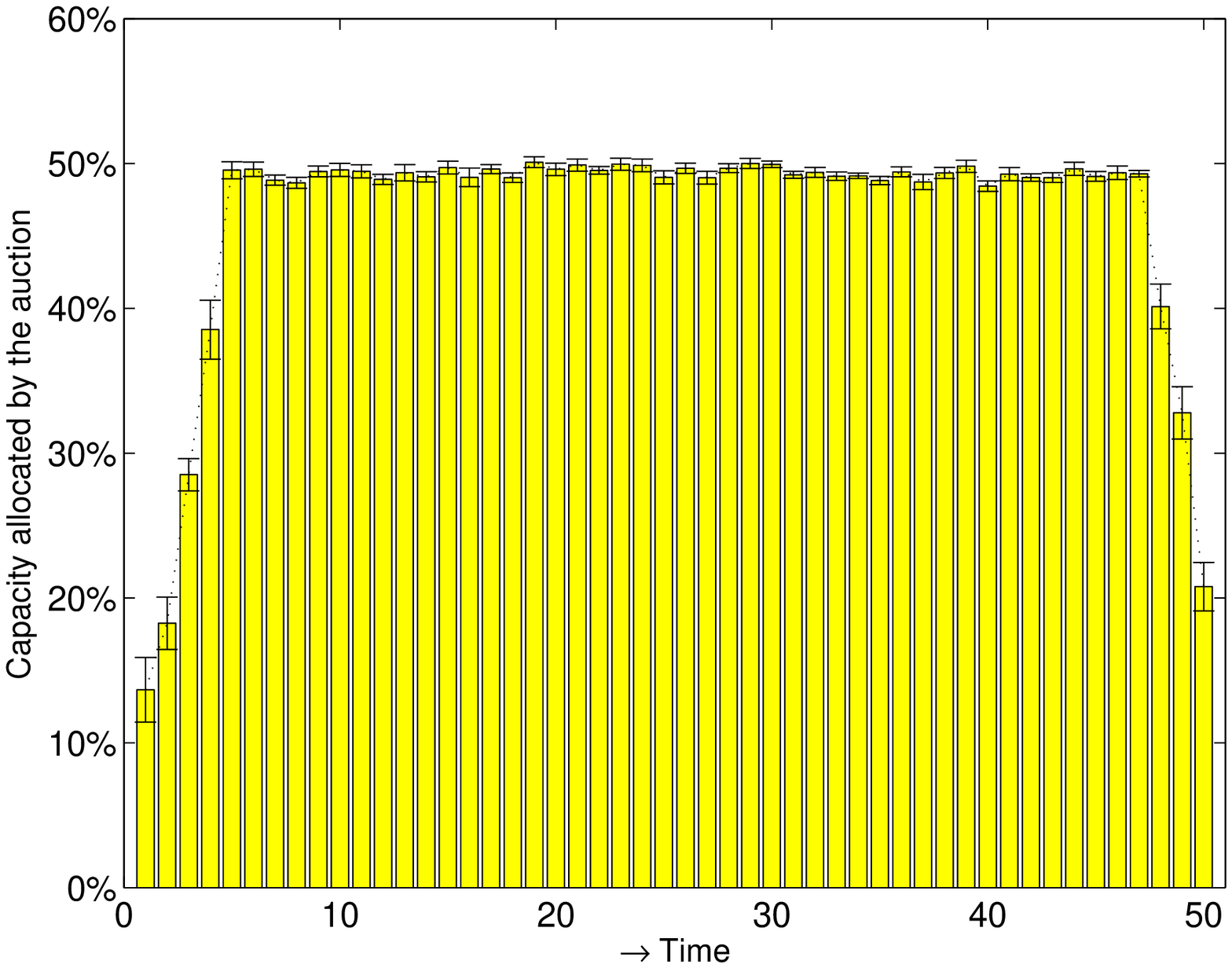}\\
(e)~~~~~~~~~~~~~~~~~~~~~~~~~~~~~~~~~~~~~~~~~~~~~~~~~~~~~~~~~~~~~(f)\\
\end{center}
\caption{Proxy phase of an auction with 50 time slots. Indices of: (a) Customer satisfaction; (b) Service mismatch; (c) Auction success; (d) Spot allocation opportunity; (e) Temporal fragmentation; (f) Capacity allocation.}
\label{ResultsFig}
\end{figure*}

Figures \ref{ResultsFig}(a)-(e) show several performance metrics including the customer satisfaction index, the service mismatch index, the auction success ratio, the spot opportunity index, the temporal fragmentation index, and the capacity allocation index. The simulation covers 50 time slots.   

The $5\%$ confidence intervals for the mean of all  performance metrics are computed for $25$ batches each one of $200$ realization of each random variable. The simulation times are $6.4$ seconds for $2,000$ runs and  $11.7$ seconds for $5,000$ runs.   The confidence intervals are rather tight; this indicates that the performance of the protocol is relatively stable for the range of parameters explored in this evaluation.

The auction success rate is high, typically above $80\%$. The initial low auction success rate is an artifact of the manner we conducted the simulation; we picked up randomly the service start up time. The spot allocation opportunity index is in turn correlated with the auction success rate and shows that  a significant fraction of the capacity is available for spot allocation. This result is correlated with the one in Figure \ref{ResultsFig}(f) which shows that on average some $50\%$ of the server capacity is not allocated by the reservation system and so is available for spot contention.

A reservation system covering $50\%$ of the server capacity is probably the most significant result; it shows that self-management based on auctions can drastically improve server utilization. We live in a world of limited resources and cloud over-provisioning is not sustainable either economically or environmentally.

The service mismatch index is fairly high, typically in the $50\%$ range and it is above  $60\%$ in a few slots. The customer satisfaction is correlated with the service mismatch and typically is in the region of $50\%$. In a realistic scenario, when coalitions maintain statistics regarding the services offered and avoid offering services unlikely to be demanded by the cloud users, the service mismatch would not affect the performance of the algorithm. Temporal fragmentation, though rather low, is undesirable. The overbidding factor  $64 ~\pm~ 2.93\%$ is another indication that the protocol needs to be fine tuned.

Self-organization cannot occur instantaneously in an adaptive system and this simple observation has important consequences. It is critical to give autonomous cloud platforms, interconnected by a hierarchy of networks,  the time to form coalitions in response to services demanded. Thus, self-management requires an effective reservation system and our results indicate that the reservation protocol is working well. 
\section{Conclusions and Future Work}
\label{Conclusions}
\medskip

Self-organization and self-management offer an appealing alternative to existing cloud resource management policies; they have the potential to significantly alter the cloud computing landscape. So far, pragmatic means for the adoption of self-organization principles for large-scale computing and communication systems have eluded us.   A main reasons for this state of affairs  is that self-management has to be coupled with some mechanisms for cooperation; these mechanisms should allow autonomous servers, to act in concert towards global system goals. Cooperation means that individual systems have to partially surrender their autonomy. Striking a balance between autonomy and cooperation is a challenging task, it requires a fresh look at the mechanics of self-organization and the practical means to achieve it.

Practical implementation of cloud self-organization is challenging for several reasons including the absence of a technically suitable definition of self-organization, a definition that could hint to practical design principles for self-organizing systems and quantitative evaluation of the results. Computer clouds exhibit the essential  aspects of complexity and it is inherently difficult to control complex systems.

We started our investigation with a realistic model of the cloud infrastructure, the hierarchical organization reported in \cite{Barroso13} which seems inherently tied to hierarchical control. First, we compared hierarchical control which based on monitoring with a market model in which the servers of a WSC place bids for service requests and found out that the latter  is much more effective than  hierarchical control \cite{Marinescu15}.  In the simple market model the servers act individually, rather than cooperating with each other, a fundamental aspect of self-organization. But cooperation is clearly needed because individual servers may not be able to supply the resources demanded by many data-intensive application. Thus, we concluded that servers have to form coalitions to offer larger pools of resources. At the same  time,  it seemed obvious to us that complex applications with multiple phases would require packages of resources offered by different coalitions. 

Algorithms for coalition formation based on combinatorial auctions are at the heart of the cloud ecosystem we propose. The path we chose seems logical as auctions have been successfully used for resource management in the past. Auctions do not require a model of the system, while traditional resource management strategies do. The auction-based protocol is scalable, and the computations can be done efficiently, though the computational algorithms involved are often fairly complex.

The results reported in Section \ref{ProtocolEvaluation} indicate that the performance of the protocol is relatively stable for the range of parameters explored in our evaluation. The protocol leads to a higher server utilization and it seems reasonable to expect that a fine-tuned version of the protocol could further improve this critical performance measure.

Two basic strategies for coalition formation are possible: (1) the one described in this paper, when coalitions are formed using information from previous combinatorial auctions, before knowing what the actual user demands are; (2)  the more natural one, when coalitions are formed in response to user demands and then combinatorial auctions are organized. The former strategy, though simple and elegant, has two obvious  drawbacks: (a) it needs the past history thus, starting from some arbitrary initial state may not work very well; (b) there may be cases when  coalitions formed based on past history are not useful to any package and, at the same time, one or more packages could benefit from the available resources of the same type, but in coalitions of different sizes. The latter strategy, creating coalitions in response to  known user requests is more complex. It requires a protocol to inform supply agents of the atrributes of desirable coalitions, a protocol for individual agents to express their willingness to join a particular coalition, and, finally, a coalition formation algorithm. Once the coalitions  are formed the bidding for service packages can take place.

These two strategies can be combined, the latter used during the initial stages, when the historic information is either missing or incomplete, and the former used when history data is available. The second drawback of the strategy described in this paper can be attenuated by allowing a second round of coalitions  formation. In this second round the unsuccessful coalitions are disassembled and new coalitions matching the needs of unsatisfied packages are formed. For example, if two coalitions of size $\theta_{1}$ and $\theta_{2}$ were unsuccessful  and a package $\mathcal{P}$  requireing a coalition of size  $\theta_{3} \le \theta_{1} + \theta_{2}$ was unsatisfied, then a second round will guarantee that $\mathcal{P}$ is satisfied and $\theta_{3}$ servers  will be reserved, instead of being left idele or offered to spot allocations.

Our future work will address this approach, as well as, other problems revealed by this investigation, e.g.,  the effects of overbidding. Overbidding is the process allowing a client to become a provisional winner of one or more service slots and then, in the final round failing to acquire some of them. This situation is critical for the first slot of an auction as the next auctions could find clients for these slots. A more difficult problem is the temporal fragmentation which does not seem to have an obvious solution.

\smallskip

{\it Dan C. Marinescu.} During the period 1984-2001 Dan Marinescu was an Associate and the Full Professor in the Computer Science Department at Purdue University in West Lafayette, Indiana. Since August 2001 he is  a Provost Professor of Computer Science at University of Central Florida. He has published more than 220 papers in referred journals and conference proceedings and several books including:  {\it Cloud Computing: Theory and Practice}, Morgan Kaufmann, 2013.

\smallskip

{\it Ashkan Paya} is a Ph.D. candidate in the Electrical Engineering and Computer Science Department at University of Central Florida pursuing his degree in Computer Science. He graduated from Sharif University of Technology in Tehran, Iran, with a BS Degree in the same major in 2011. His research interests are in the area of resource management in large-scale systems and cloud computing.

\smallskip

{\it John Morrison} is the founder and director of the Centre for Unified Computing. He is a co-founder and director of the Boole Centre for Research in Informatics, a principle investigator in the Irish Centre for Cloud Computing and Commerce and a co-founder and co-director of Grid-Ireland. Prof. Morrison has held a Science Foundation of Ireland Investigator award and has published widely in the field of Parallel Distributed and Grid Computing. He is a principle investigator in the Irish Centre from Cloud Computing and Commerce, where he leads the Service LifeCycle Group. He has been the guest editor on many journals
including the Journal of Super Computing and the Journal of Scientific Computing. He is on the Editorial Board of Multi-Agent and Grid Systems: An International Journal, published by ISO Press, and the International Journal of Computational Intelligence: Theory and Practice (IJCITP).
He is a senior member of the ACM and a senior member of the IEEE. 


\begin{thebibliography}{11}

\bibitem{Ausubel02}
L.~Ausubel and P. R. Milgrom.
\newblock ``Ascending auctions with package bidding.''
\newblock {\it Frontiers of Theoretical Economics}, {\bf 1}(1):1--42, 2002.

\bibitem{Ausubel04}
L. M. ~Ausubel and P.~Cramton.
\newblock ``Auctioning many divisible goods.''
\newblock {\it Journal European Economic Assoc.}, {\bf 2}(2-3):480�-493, 2004.

\bibitem{Ausubel06}
L.~Ausubel, P.~Cramton, and P.~Milgrom.
\newblock ``The clock-proxy auction: a practical combinatorial auction design.''
\newblock {\it Chapter 5,} in {\it Combinatorial Auctions},
P. Cramton, Y. Shoham, and R. Steinberg, Eds. MIT Press, 2006.

\bibitem{Barroso07}
L. A.~Barroso and U.~H\"ozle.
\newblock ``The case for energy-proportional computing.''
\newblock {\it IEEE Computer,} {\bf 40}(12):33--37, 2007.

\bibitem{Barroso13}
L. A. Barossso, J. Clidaras, and U.H\"ozle.
\newblock {\it The Datacenter as a Computer; an Introduction to the Design of Warehouse-Scale Machines.} (Second Edition).
\newblock Morgan \& Claypool, 2013.

\bibitem{Blackburn10}
M. Blackburn and A. Hawkins.
\newblock ``Unused server \break  survey results analysis.''
\newblock {\it www.thegreengrid.org/\break media/WhitePapers/Unused  \%20Server\%20Study \break \_WP\_101910\_v1.
ashx?lang=en} (Accessed on December 6, 2013).

\bibitem{Bruneo14}
D. Bruneo.
\newblock  ``A stochastic model to investigate data center performance and QoS in IAAS cloud computing systems.''
\newblock {\it IEEE Trans. on Parallel and Distributed Systems,}  {\bf 25}(3):560--569, 2014.

\bibitem{Carroll10}
T. E. Carroll and D. Grosu.
\newblock ``Formation of virtual organizations in grids: a game-theoretic approach.''
\newblock {\it Concurrency and Computation: Practice and Experience,} {\bf 22}(14):1972--1989, 2010.

\bibitem{Chaisiri12}
S. Chaisiri, B. Lee, and D. Niyato.
\newblock  ``Optimization of resource provisioning cost in cloud computing.''
\newblock {\it IEEE Trans. on Services Computing,} {\bf  5}(2):164--177, 2012.

\bibitem{Chang10}
V.~Chang, G.~Wills, and D.~De Roure.
\newblock ``A review of cloud business models and sustainability.''
\newblock {\it Proc. IEEE 3rd Int. Conf. on Cloud Computing}, pp. 43--50, 2010.


\bibitem{Clarke71}
E. H. Clarke.
\newblock ``Multipart Pricing of Public Goods.''
\newblock {\it Public Choice}, {\bf IX}:13--33, 1971.

\bibitem{Groves73}
T. Groves.
\newblock ``Incentives in teams.''
\newblock {\it Econometrica,} {\bf 41}:617--631, 1973.

\bibitem{Gutierrez10}
J.~O.~Gutierrez-Garcia and K.- M.~Sim.
\newblock ``Self-organizing agents for service composition in cloud computing.''
\newblock {\it Proc IEEE 2nd Int. Conf. on Cloud Computing Technology and Science,}
pp. 59--66, 2010.

\bibitem{He05}
L. He and T. R. Ioerger.
\newblock ``Forming resource-sharing coalitions: a distributed resource allocation mechanism for self-interested agents in computational grids.''
\newblock {\it Proc. ACM Symp. on Applied Computing,} pp. 84--91, 2005.

\bibitem{Khan09}
S. U. Khan and I. Ahmad.
\newblock ``A cooperative game theoretical technique for joint optimization of energy 
consumption and response time in computational grids.''
\newblock {\it IEEE Trans. on Parallel and Distributed Systems,} {\bf 20}(3): 346-360, 2009.

\bibitem{Lerman00}
K. Lerman and O. Shehory.
\newblock ``Coalition formation for large-scale electronic markets.''
{\it Proc. ICMAS 2000 - 4th Int. Conf on Multiagent Systems,} pp. 167--174, 2000.

\bibitem{Li02}
C. Li and K. Sycara.
\newblock ``Algorithm for combinatorial coalition formation and payoff division in an electronic marketplace.''
\newblock {\it Proc. AAMAS02 - First Joint Int. Conf. on Autonomous Agents and Multiagent Systems,} pp. 120--127, 2002.

\bibitem{Li13}
H. Li, C. Wu, Z. Li, and F. Lau.
\newblock ``Profit-maximizing virtual machine trading in a federation of selfish clouds.''
\newblock {\it Proc. of the IEEE INFOCOM,} pp. 25--29, 2013.

\bibitem{Lim09}
H C.~Lim,  S.~Babu, J. S.~Chase, and S. S.~Parekh.
\newblock ``Automated control in cloud computing: challenges and opportunities.''
\newblock {\it Proc. First Workshop on Automated Control for Datacenters and Clouds,},
ACM Press, pp. 13--18, 2009.

\bibitem{Marinescu10}
D. C.~Marinescu,  C.~Yu, and G. M.~Marinescu.
\newblock {``Scale-free, self-organizing very large sensor networks.''} Journal of Parallel and Distributed Computing (JPDC), {\bf 50}(5):612--622, 2010.

\bibitem{Marinescu13}
D. C. Marinescu.
\newblock {\it Cloud Computing; Theory and Practice.}
\newblock Morgan Kaufmann, a division of Elsevier, Amsterdam, New York,  2013.

\bibitem{Marinescu15}
D. C. Marinescu, A. Paya, J. P. Morrison, and P. Healy.
\newblock ``Distributed hierarchical control versus an economic model for cloud 
resource management.''
\newblock {\it http://arxiv.org/pdf/1503.01061v1.pdf}, March 2015.

\bibitem{Mashayekhy15}
L.Mashayekhy, M.M. Nejad, and D. Grosu.
\newblock ``Cloud federations in the sky: formation game and mechanisms.''
\newblock {\it IEEE Trans. on Cloud Computing,}  2015 (to appear).

\bibitem{Muller06}
I. M\"uller, R. Kowalczyk, and P. Braun.
\newblock ``Towards agent-based coalition formation for service composition.''
\newblock {\it  Proc. IEEE/WIC/ACM Int. Conf. on Intelligent Agent Technology,} pp. 73-80, 2006.

\bibitem{Niyato11}
D.Niyato, A.Vasilakos, and Z.Kun.
\newblock ``Resource and revenue sharing with coalition formation of cloud providers: Game theoretic approach.''
\newblock {\it Proc. IEEE/ACM Intl. Symp. on Cluster, Cloud and Grid Comp.,} pp. 215--224, 2011.

\bibitem{Paya15}
A. Paya and D. C. Marinescu.
\newblock ``Energy-aware load balancing and application scaling for the cloud ecosystem.''
\newblock {\it IEEE Trans. on Cloud Computing.}  doi:10.1109/TCC.2015.2396059.

\bibitem{Penmatsa06}
S. Penmatsa and A. T. Chronopoulos.
\newblock ``Price-based user-optimal job allocation scheme for grid systems.''
\newblock  {\it Proc of Parallel and Distributed Processing Symposium,}  pp. 8-16, April 2006.

\bibitem{Rahwan09}
T. Rahwan, S. D. Ramchurn, N. R. Jennings, and A. Giovannucci. 
\newblock ``An anytime algorithm for optimal coalition structure generation.''
\newblock {\it Journal of Artificial Intelligence Research,} {\bf 34}:521--567, 2009.

\bibitem{Samaan14}
N. Samaan.
\newblock ``A novel economic sharing model in a federation of selfish cloud providers.''
\newblock {\it IEEE Trans. on Parallel and Distributed Systems,} {\bf 25}(1):12--21, 2014.

\bibitem{Sandholm99}
T. W. Sandholm, K. S. Larson, M. Andersson, O. Shehory, and F. Tohm�. 
\newblock ``Coalition structure generation with worst case guarantees.''
\newblock {\it Artificial Intelligence,} {\bf 111}(1-2):209--238, 1999.

\bibitem{Sen00}
S. Sen and P. S. Dutta.
\newblock ``Searching for optimal coalition structures.''
\newblock {\it Proc. ICMAS 2000 - 4th Int. Conf on Multiagent Systems,} pp. 287--295, 2000.


\bibitem{Sims03}
M. Sims, C. V. Goldman, and V. Lesser.
\newblock ``Self-organization through bottom-up coalition formation.'' 
\newblock {\it Proc. Int. Conf. on Autonomous Agents and Multi Agent Systems,} pp. 867--874, 2003.


\bibitem{Subrata08}
R. Subrata, A. Y. Zomaya, and B. Landfeldt.
\newblock ``Game-theoretic approach for load balancing in computational grids.'' 
\newblock {\it IEEE Trans. on Parallel and Distributed Systems,} {\bf 19}(1):66--76, 2008.

\bibitem{deVries03}
S de Vries and R. Vohra.
\newblock ``Combinatorial auctions; a survey.''
\newblock {\it INFORMS Journal of Computing}, {\bf 15}(3):284--309, 2003.

\bibitem{Zhang03}
H-J. Zhang, Q-H. Li, and Y-L. Ruan.
\newblock ``Resource co-allocation via agent-based coalition formation in computational grids.''
\newblock {\it Proc Second Int. Conf. on Machine Learning and Cybernetics,}, pp. 1936--1940, 2003.

\bibitem{Wei10}
G. Wei, A. Vasilakos, Y. Zheng, and N. Xiong.
\newblock ``A game-theoretic method of fair resource allocation for cloud computing services.'' 
\newblock {\it The Journal of Supercomputing,} {\bf 54}(2):252--269, 2010.


\end{thebibliography}
\end{document}